\documentclass[twocolumn,showpacs,prl,superscriptaddress,article]{revtex4-1}
\usepackage[english]{babel}
\usepackage{amsmath,amssymb}
\usepackage{amsfonts}
\usepackage{mathtools}
\usepackage{siunitx}
\usepackage{libertine}
\usepackage[usenames,dvipsnames]{color}
\usepackage{hyperref}
\usepackage{blindtext}
\hypersetup{colorlinks}
\hypersetup{linkcolor=black, citecolor=black, urlcolor=blue}
\usepackage[utf8]{inputenc}
\usepackage{color}
\usepackage{subcaption}
\usepackage[normalem]{ulem}
\usepackage{xcolor}
\usepackage{graphicx}

\newcommand{\dif}{\mathrm{d}}
\begin{document}
\title{Flow-induced surface charge heterogeneity in electrokinetics due to Stern-layer conductance coupled to reaction kinetics}

\author{B. L. Werkhoven}
\affiliation{Institute for Theoretical Physics, Center for Extreme Matter and Emergent Phenomena, Utrecht University, Princetonplein 5, 3584 CC, Utrecht, The Netherlands}
\author{J.C. Everts}
\affiliation{Department of Physics, Faculty of Mathematics and Physics, University of Ljubljana, Jadranska 19, 1000 Ljubljana, Slovenia}
\author{S. Samin}
\affiliation{Institute for Theoretical Physics, Center for Extreme Matter and Emergent Phenomena, Utrecht University, Princetonplein 5, 3584 CC, Utrecht, The Netherlands}
\author{R. van Roij}
\affiliation{Institute for Theoretical Physics, Center for Extreme Matter and Emergent Phenomena, Utrecht University, Princetonplein 5, 3584 CC, Utrecht, The Netherlands}


\begin{abstract}
We theoretically study the electrokinetic problem of a pressure-induced liquid flow through a narrow long channel with charged walls, going beyond the classical Helmholtz-Schmolukowski picture by considering the surprisingly strong combined effect of (i) Stern layer conductance  and (ii) dynamic charge-regulating rather than fixed surface charges. We find that the water flow induces, apart from the well-known streaming potential, also a strongly heterogeneous surface charge and zeta-potential on chemically homogeneous channel walls. Moreover, we identify a novel steady state with a nontrivial 3D electric flux with 2D surface charges acting as sources and sinks. For a pulsed pressure drop our findings also provide a first-principles explanation for ill-understood experiments on the effect of flow on interfacial chemistry [D. Lis {\em et al.}, Science {\bf 344}, 1138 (2014)].    
\end{abstract}

\maketitle

\begin{figure}[!ht]
\centering
\includegraphics[width=0.5\textwidth]{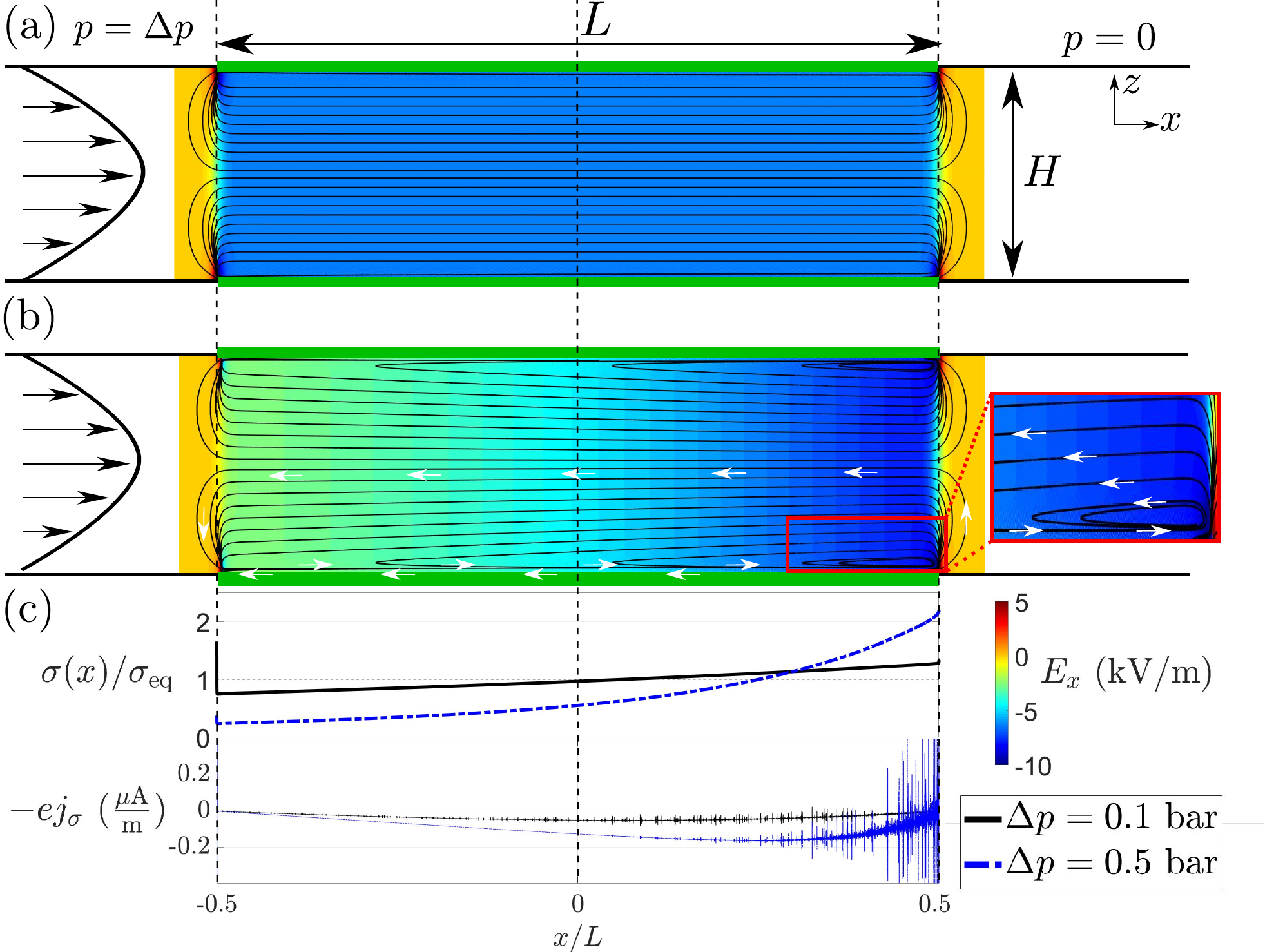}
\caption{Streamlines of the net charge flux and colour map of the tangential electric field $E_x$ near the charged surfaces (green stripes) of a rectangular channel with a pressure drop $\Delta p=0.5$ bar between in- and outlet at $x=\pm L$, (a) with vanishing Stern-layer conduction ($D_s=0$) resulting in a fixed surface charge of $-e\sigma_{\rm eq}=-0.069~e$/nm$^2$ that mimics silica at pH=6.5, and in (b) with non-zero Stern-layer conductance and our dynamic charge regulation model.  (c) Flow-induced heterogeneous surface charge density $\sigma(x)$ and surface charge flux $-ej_{\sigma}$ for $\Delta p=0.1, 0.5$ bar for the parameters of case of (b).}
\label{fig:ss}
\end{figure}

The flow of water along a solid surface such as glass, rock or an electrode is of profound interest in fields as diverse a geosciences (rivers, erosion) \cite{geology}, oil-field engineering (enhanced oil recovery) \cite{reservoirrock}, and micro- and nano-fluidics. The Poiseuille flow through a long channel due to a pressure drop between in- and outlet is a textbook example, in which the stationary Navier-Stokes equation with no-slip boundary conditions on the channel surface gives rise to a parabolic flow profile (represented in Fig. \ref{fig:ss}) that is proportional to the pressure drop. In many cases relevant for e.g. microfluidics and blue-energy harvesting \cite{blauw1,blauw2}, however, a liquid flow induces a much richer phenomenology, often due to surface charges on the channel walls that interact with the ionic species in the liquid. In such a channel an applied pressure drop does not only induce a fluid flow but also a net electric current due to advection of the so-called electric double layer (EDL), which is the diffuse layer of mobile ions that screen the electrode in the nanometer vicinity of the charged surface.  In closed-circuit conditions this so-called ``streaming current" can persist in a stationary state, but in open-circuit conditions it leads to the build-up of net charge and hence a potential difference between the outlet and the inlet of the channel, the so-called ``streaming potential" $\Delta\Phi_S$ derived long ago by Helmholtz \cite{helmholtz} and Smoluchowski  \cite{smoluchowski},  as 
\begin{equation}\label{eq:smsurf}
\Delta \Phi_S=\dfrac{-\zeta \epsilon}{\eta G}\Delta p.
\end{equation}
Here $\zeta$ is the (zeta-)potential at the slipping planes, $\epsilon$ and $\eta$ the dielectric permittivity and the shear viscosity of the liquid, respectively, and $\Delta p$ the pressure drop that drives the Poiseuille flow. The total channel conductivity $G=G_b+2G_s/H$ of a channel of height $H$ is well known to consist not only of a bulk contribution $G_b$ but also of two surface contributions $G_s/H$ to account for conduction processes close to the channel surfaces  \cite{lyklemaboek}. 
The dimensionless Duhkin number Du=$G_s/G_bH$ characterises the relative importance of the surface-to-bulk conduction \cite{lyklemaboek}. It is important to realise that $G_s=G_s^d+G_s^S$ not only contains a contribution $G_s^d$ from the relatively high density of charge carriers in the diffuse part of the EDL, as first recognised by Bikerman in 1933 \cite{bikerman1,bikerman2}, but {\em also} a contribution $G_s^S$ from the quasi-2D Stern layer where the surface charges reside \cite{overbeekkruyt}.  In fact, a substantial body of literature exists that not only confirms the finite charge mobility in the Stern layer for different types of (insulating) materials such as PMMA, silica or clay \cite{lyklemareview,saini, lobbus,sonnefeldsilica, obrianclay, leroyclay}, but even that the in-plane charge mobility is comparable to the mobility of simple ions in bulk electrolytes \cite{minormob,lobbusmob,lyklemareview}. The lateral conductance in the Stern layer is the first key ingredient of this Letter.

Eq.(\ref{eq:smsurf}) stems from a linear-response analysis, in which the prefactor $-\zeta\epsilon/\eta G$ is assumed to be a constant for a given channel and transported fluid. Motivated by inherently heterogenoeus biological surfaces and by microfluidic applications with patterned electrodes, extensions towards periodic \cite{ajdari} and steplike \cite{khairsquires} variations of $\zeta$ were considered. Heterogeneity of $\zeta$ not only leads to normal components of the ionic fluxes \cite{ajdari,khairsquires}, but also to the notion of the so-called healing length $\ell=G_s/G_b=H\mbox{Du}$ as the governing lateral length scale \cite{khairsquires}. However, in this Letter we will for the first time show that defect-free and unpatterned surfaces, charged over a finite length, can exhibit flow-induced heterogeneities with the surface charge density and the zeta-potential varying over the full length of the charged surface, even if Du$\ll 1$. Here one should realise that most surfaces in contact with water do not have a fixed charge but obtain their net charge by {\em regulation} processes, in which for instance a fraction $f$ of the neutral surface groups SC disscociates into a covalently bound negatively charged surface group S$^-$ and a released cation C$^+$. The reaction SC$\rightleftharpoons$S$^-$+C$^+$ is characterised by an equilibrium constant $K$, which together with the C$^+$ concentration at the surface $\rho_{C,s}$ determines the equilibrium Langmuir desorption isotherm $f=(1+\rho_{C,s}/K)^{-1}$ \cite{chargeregulation,CRChan,CRBorkovec}. Although the importance of charge regulation was indeed recognised in earlier works on the electrophoresis of colloidal particles, the underlying equilibrium Langmuir desorption isotherm has so far always been assumed \cite{zukoski,mangelsdorf}. In this Letter we will introduce {\em out-of-equilibrium} charge regulation as a second key ingredient, in which the rates of adsorption ($k^{\rm ads}$) and desorption ($k^{\rm des}$) play a key role individually rather than only their ratio $K=k^{\rm des}/k^{\rm ads}$.  In fact, by tuning the chemical rates to the reaction-limited regime, we will see that our theory provides a natural first-principles explantion for puzzling recent experiments that show a profound influence of a fluid flow on the interfacial chemistry \cite{lis}, provided Stern layer conduction and out-of-equilibrium charge regulation are taken into account simultaneously.  We expect that this intricate interplay between dynamic charge regulation and Stern layer conduction will play an equally important role in many nanoflow problems of recent interest \cite{blauw2,nano1,nano2,nano3}.

The system we consider in this Letter, sketched in Fig \ref{fig:ss}, consists of two bulk aqueous reservoirs connected by a wide rectangular channel of length $2L$ and height $H$, with lateral and normal Cartesian coordinates $x\in[-L,L]$, $z\in[0,H]$, and with translational invariance in the lateral $y$ direction. The reservoirs contain three monovalent ionic species labelled by $i=+,-,C$ with valency $z_+=z_C=-z_-=1$ and with bulk concentrations $\rho_{i,b}$, satisfying neutrality $\sum_iz_i\rho_{i,b}=0$. The Debye screening length is given by $\lambda_D=\sqrt{\epsilon k_{\rm B}T/e^2\sum_i z_i^2\rho_{i,b}}$, with $k_{\rm B}$ the Boltzmann constant, $T$ the temperature, and $e$ the proton charge. 

We denote the time- and position dependent ionic density profiles and fluxes (actually flux densities) by $\rho_i({\bf r},t)$ and ${\bf J}_i({\bf r},t)$, respectively, the electric potential by $\psi({\bf r},t)$, the (identical) surface charge density in the planes $z=0$ and $z=H$ by $-e\sigma(x,t)$, and the fluid velocity profile by ${\bf u} ({\bf r},t)$. The ion flux ${\bf J}_i$  is composed of diffusive, conductive, and advective contributions, and is related to $\partial_t\rho_i$ by the continuity equation. The Poisson equation accounts for Coulomb interactions, and the incompressible Navier-Stokes equation, including an electric body force and a pressure gradient $\nabla p$, describes the fluid flow. Collecting all this we obtain the well-known Poisson-Nernst-Planck-Navier-Stokes (PNPNS) equations \cite{hunter} with Gauss' law and no-slip boundary conditions,
\begin{gather}
\begin{aligned}
&\dfrac{\partial \rho_i}{\partial t}=-\nabla\cdot  {\bf J}_i; \hspace{3mm} {\bf J}_i =-D\left(\nabla\rho_i+\frac{ez_i\rho_i}{k_{\rm B}T}\nabla\psi\right)+\rho_i{\bf u};\\
&m\dfrac{\partial {\bf u}}{\partial t}=-m({\bf u} \cdot\nabla){\bf u}-\nabla p+\eta\nabla^2{\bf u}-\sum\limits_i z_ie\rho_i \nabla \psi;\\
&\nabla\cdot {\bf u}=0; \qquad \nabla^2\psi=-\frac{e}{\epsilon}\sum\limits_iz_i\rho_i;\\
&{\bf u}_s=0; \hspace{11mm} {\bf n}_s\cdot \nabla \psi_s=\frac{e\sigma}{\epsilon}.
\end{aligned}
\label{eq:equations}
\raisetag{50pt}
\end{gather}
Here $D$ is the diffusion constant, assumed to be equal for all ion species, and $m$ is the mass density of water.  Throughout this Letter the subscript ``s" denotes a surface quantity, e.g. ${\bf u}_s$ is the fluid velocity at the surface and ${\bf n}_s$ is the unit surface normal pointing into the water. 

The standard PNPNS equations (\ref{eq:equations}) are to be contrasted with the novel boundary conditions that we impose in this Letter, where we describe regulation of the surface charge $\sigma$ combined with Stern layer conductance \cite{SIeqs}. The bottom and the top surfaces  each contain an identical, chemically homogeneous patch for $x\in[-L/2,L/2]$ that can become negatively charged due to the desorption reaction SC$\rightleftharpoons$S$^-+$ C$^+$. Therefore, a non-vanishing cationic counterion flux $-{\bf n}_s\cdot{\bf J}_{C,s}(x)$ entering the surface is possible, whereas the other two ionic species $\pm$ are chemically inert and satisfy the no-flux boundary conditions ${{\bf n}_s\cdot\bf J}_{\pm,s}=0$. Within the Stern layer we introduce the lateral (surface) current  $-ej_{\sigma}(x)$, satisfying the continuity equation for the surface charge density as  
\begin{equation}
\frac{\partial \sigma}{\partial t}=- \frac{\partial j_{\sigma}}{\partial x}+{\bf n_s\cdot J_{\rm C,s}},
\label{eq:continuity}
\end{equation}
which explicitly couples the 3D flux of cations as a source term for the 2D surface density.
We describe the net flux of C$^+$ towards the surface in terms of simple reaction kinetics with an adsorption flux $k^{\rm ads}\sigma\rho_{C,s}$ and a desorption flux $k^{\rm des}(\Gamma-\sigma)$, with $\Gamma$ the total number of chargeable sites per unit area. Additionally, we assume a Nernst-Planck like equation for $j_{\sigma}$, with a diffussive and a conduction contribution, where the former is modified to account for forbidden multiple ad- and desorption on a single site \cite{lattice}:
\begin{eqnarray}
j_{\sigma}(x)&=&-D_s\left(\frac{1}{1-\sigma/\Gamma}\frac{\partial\sigma}{\partial x}-\frac{e\sigma}{k_{\rm B}T}\frac{\partial \psi_s}{\partial x}\right);\label{eq:bc1}\\
-{\bf n_s\cdot J_{\rm C,s}}&=&-k^{\rm des}(\Gamma-\sigma)+k^{\rm ads}\sigma\rho_{\rm C,s},\label{eq:bc2}
\end{eqnarray}
where $D_s$ the surface diffusion constant, which we have seen to be comparable to the bulk diffusion coefficient $D$.  If we impose static equilibrium conditions, in particular ${\bf J}_C= 0$, Eqs. (\ref{eq:continuity}) - (\ref{eq:bc2}) reduce to the standard Langmuir desorption isotherm where $\sigma/\Gamma$ equals the fraction $f$ of charged sites introduced above \cite{SIequilibrium}. In the case of a pressure-induced flow, however, the streaming potential generates an in-plane electric field component $\partial_x \psi_s$, which according to Eq.(\ref{eq:bc1}) not only drives a finite $j_{\sigma}$ if $D_s\neq0$, but for a charge-regulating surface {\em also} a finite ${\bf n}_s\cdot{\bf J}_{\rm C,s}$ and a surface heterogeneity $\partial_x\sigma$ according to Eqs. (\ref{eq:continuity}) and (\ref{eq:bc2}). As a consequence the zeta-potential $\zeta(x)=\psi(x,0)-\psi(x,H/2)$ becomes heterogeneous too, and hence a nontrivial self-consistency problem emerges in which the streaming potential not only determines $\zeta(x)$ but also depends on it (see e.g. Eq.(\ref{eq:smsurf})). Interestingly, this flow-induced surface heterogeneity does {\em not} require relatively narrow channels or high Du.

We solve the set of non-linear equations (\ref{eq:equations})-(\ref{eq:bc2}) numerically using the Finite-Elements software COMSOL Multiphysics. For computational reasons we take at each side of the chargeable surface an uncharged patch of length $L/2$ to allow entrance and exit effects on the fluid flow driven by a  pressure drop $\Delta p$ to essentially die out \cite{SICOMSOL}. Due to the crucial role played by the chemical reaction we must fully resolve the EDL in order to accurately determine $\rho_{C,s}$. The thin-EDL approximation \cite{hunter} is therefore not possible here. In this Letter we choose parameters that represent silica at pH=6.5, such that $-\log_{10}\rho_{\rm C,b}({\rm M})=6.5$, $\Gamma=4.6$~nm$^{-2}$, and ${\rm p}K=6.75$ (an average over the widely varying reported values \cite{silica,silica2,silica3}), with millimolar added salt concentrations $\rho_{\pm,\rm b}\simeq1$~mM such that $\lambda_D=10$~nm. The single reaction mechanism assumed here is actually too simple to capture the behaviour of silica quantitatively, but it serves our purposes here as a generic case. Under these conditions, the equilibrium surface charge and potential are $-e\sigma_{\rm eq}=-0.069~e$/nm$^2$ and $\zeta_{eq}=-93$~mV. Throughout we set $D=10^{-9}~$m$^2/$s such that $G_b=7.5$ mS/m and $G_s^d\approx 1.2$ nS \cite{Delgado}. In agreement with Stern-layer mobilities discussed above, we either set $D_s=D$ or $D_s=0$ to study presence or absence of Stern-layer conductance, respectively. We furthermore focus on a channel height $H=1\,\mu$m, i.e. $H\gg\lambda_D$ and Du $\simeq 0.16$. Apart from the channel length $L$ the only remaining system parameter is the time scale of the adsorption-desorption process, which will be fitted to experiments below. For computational efficiency we set $k^{\rm des}=2\times10^{-4}$~s$^{-1}$ for now, which is comparable to certain photocatalytic rates \cite{rates} and comfortably in the reaction-limited regime as we will see. 

In Fig. \ref{fig:ss} we show the steady-state field lines of the ionic charge flux ${\bf J}_e= \sum_i z_i {\bf J_i}$ and a colourmap of the $x$-component of the electric streaming field $E_x$  for a channel of height $H=1\mu{\rm m}$ and total length $2L=60 \mu$m, and a pressure drop $\Delta p=0.5$ bar, in (a) without Stern layer conduction ($D_s=0$), and in (b) in the presence of both Stern layer conduction ($D_s=D$) and charge regulation. The resulting maximum fluid velocity is approximately 0.1 m/s, three orders of magnitude higher than the elecotro-osmotic slip velocity induced by the electric field, i.e. the body forces (last term NS Eq. (\ref{eq:equations})) are negligible \cite{SIBF}. A striking difference between (a) and (b) are non-parallel field lines in (b), even far outside the EDL, and a much weaker electric field especially for $x\in[-L/2,0]$ in (b).  We can trace these two features back to a nonzero surface current $j_{\sigma}(x)$ and a strong heterogeneity of the surface charge profile $\sigma(x)$; both extend over the full width $L$ as shown in Fig. \ref{fig:ss}(c).  This shows that, in addition to the inherent heterogeneities of silica in equilibirum conditions \cite{roke}, surfaces can exhibit dynamical heterogeneities.  We note that the diffusive and conductive contributions to $j_{\sigma}$ (see Eq. (\ref{eq:bc1})) are counteracting and individually three orders of magnitude larger than $j_{\sigma}$, {\em i.e.} both are essential to obtain this steady state. The near-cancellation is the cause of the numerical noise observed for $j_{\sigma}$, and furthermore leads to the suprising conclusion that the effects persist even for Du$\ll 1$ \cite{SIDuhkin}.  Fig. \ref{fig:ss}(b) also shows that ${\bf J}_e$ and $E_x$ depend not only on $z$ but also on $x$, even far outside the EDL. Note that a lateral heterogeneous charge current has also been reported in the case of a (highly conducting) metallic surface \cite{duval}. 

For $\Delta p=0.1$~bar the heterogeneous profile $\sigma(x)$ shown in Fig. \ref{fig:ss}(c) is essentially linear in $x$, locally lower/higher by about $\pm 25$\% of $\sigma_{eq}$ at the inlet/outlet side of the chargeable area. For $\Delta p=0.5$ bar, however, $\sigma(x)$ is strongly nonlinear with deviations ranging from $-75\%$ to as high as $+ 100\%$ from $\sigma_{eq}$ at the edges.  In equilibrium, such a change in the surface charge would correspond to a pH varying between 4.9 and 7.4, i.e.  concentrations of C$^+$ that are a factor of 10 higher and lower. The laterally averaged charge in this case decreases to a value as low as $\langle\sigma\rangle=0.7\sigma_{eq}$. Additionally, $\langle \zeta\rangle$ also decreases, which causes a breakdown of Eq. (\ref{eq:smsurf}) \cite{SISmol}. Therefore the local as well as the average surface charge are not at all (quasi-)static quantities, but fully dynamic properties of the solid-fluid interface that can be tuned by the fluid flow in the channel. The sharp peaks of $\sigma$ at $x\simeq\pm L/2$ in Fig. \ref{fig:ss}(c) are expected in a range of $\lambda_D$ next to an uncharged area \cite{boon}. 

We can identify four different time scales that govern the dynamics of this system, (i) the EDL diffusion time $\tau_{\rm dif}=\lambda_D^2/D$ which is only about 100 ns for our parameter choice; (ii) the advection time $\tau_{\rm adv}=L/u_x(\lambda_D)$ for an ion in the EDL to be advected parallel to the surface over a distance $L$, of the order of ms in all cases studied here; (iii) the conduction time $\tau_{\rm cond}=L\sigma_{eq}/j_{\sigma}$ for a charge in the Stern layer to traverse a lateral distance $L$, which is of the order of seconds here; and (iv) the chemical reaction time  $\tau_{\rm reac}=(k^{\rm ads}\rho_{C,s})^{-1}$ \cite{SItreac} of the order of an hour here. We found that significant heterogeneities only occur if $\tau_{\rm reac}$ exceeds the three others, i.e. if the system is in the reaction-limited rather than in the diffusion-, advection- or conduction-limited regime. This can be qualitatively understood, e.g. if $\tau_{\rm reac}\ll\tau_{\rm cond}$ chemical equilibration would take place before any conductive flux can develop. Note also that $\tau_{\rm cond}\propto D_s^{-1}$ confirms the crucial role played by a finite surface conduction, since $D_s=0$ would cause the system to be conduction- rather than reaction-limited. As long as this ordering of time-scales is obeyed and $D_s/D=\mathcal{O}(1)$, as noted already on the basis of Refs. \cite{minormob,lobbusmob,lyklemareview}  the exact value of $D_s$ has no significant effect on the presented results.

\begin{figure}[!ht]
\centering
\includegraphics[width=0.45\textwidth]{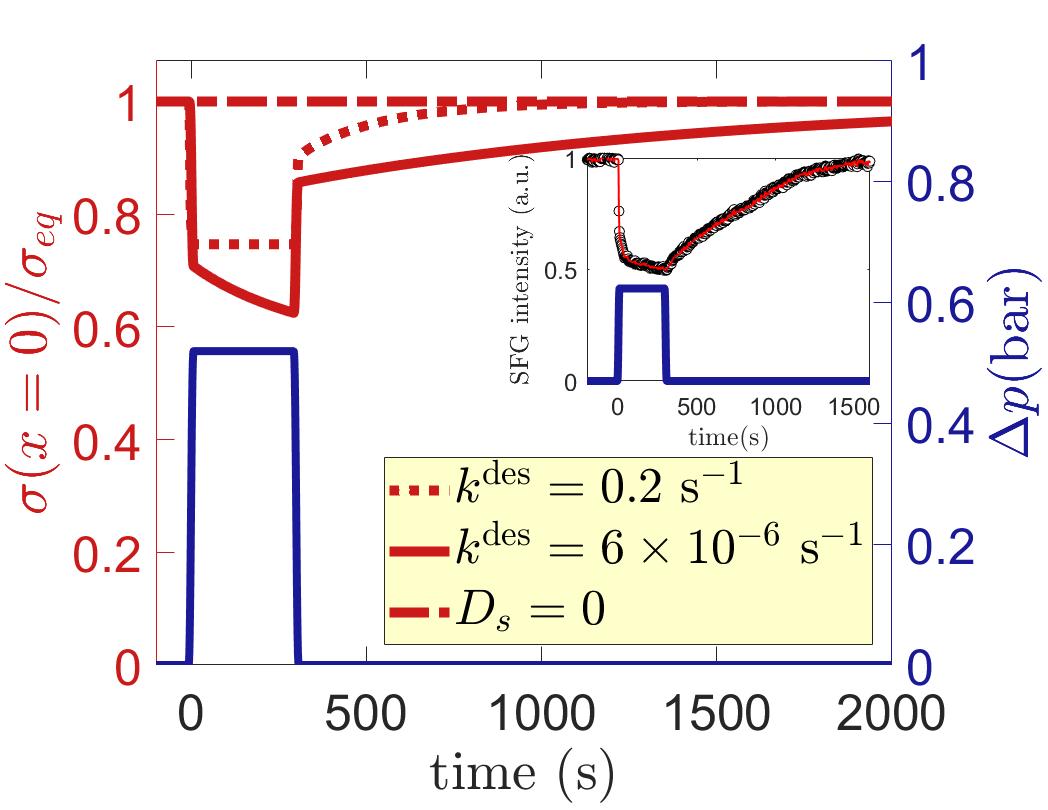}
\caption{Time-dependent pressure drop $\Delta p(t)$ (blue) in a channel of dimensions $H=1\mu$m and $L=40\mu$m, together with the resulting surface charge $\sigma(x=0,t)$ in the middle of the channel, for a silica surface at pH=6.5 (see text) with desorption rate $k^{\rm des}=6\times10^{-6}~$s$^{-1}$ ($\tau_{\rm reac}=1.7\times 10^{3}$~s) (red), to be compared with the experimental data of \cite{lis} shown in the inset. The red dotted line shows the case of a non-conducting Stern layer with $D_s=0$, and the dashed line the case with desorption rate $k^{\rm des}=0.2~$s$^{-1}$ ($\tau_{\rm reac}=0.05~$s).}
\label{fig:sc}
\end{figure}

So far we have seen that the stationary state of a charge-regulating and conducting surface exposed to a fluid flow becomes heterogeneously charged in a stationary state. In an exciting experiment in 2014, however, the full relaxation dynamics of the surface charge of silica upon an applied water pressure {\em pulse} was measured in an experiment that combines microfluidics and Sum Frequency Generation (SFG) \cite{lis}, albeit only at the central position (here $x=0$) in the channel. By ruling out alternative interpretations the authors of \cite{lis} attribute their time-dependent SFG-signal to a time-dependent surface charge $\sigma(x=0,t)$.  Here we confirm this interpretation by showing that our theory provides a microscopic explanation for the time dependence of the surface charge, which in the experimements (see inset Fig. \ref{fig:sc} or Fig. 2D of \cite{lis}) consists of a quasi-instantaneous initial reduction by 40\% (on the time scale of seconds) upon switching on the flow followed by a further reduction by an additional 10\% on the time scale of minutes, and upon switching off the flow a very slow relaxation (on the time scale of tens of minutes) back to equilibrium. In Fig \ref{fig:sc} we show a time-dependent pressure pulse (blue) similar to the experimental one as well as the surface charge density  $\sigma(x=0,t)$ (red) that follows from our theory. Here we use the same silica parameters and bulk concentrations as before in Fig. \ref{fig:ss}(b), again at pH=6.5 but now with the desorption rate $k^{\rm des}=6\times 10^{-6}~$s$^{-1}$ as the only ``fit"-parameter. This corresponds to $\tau_{\rm reac}=1.7\times 10^{3}~$s, which sets the transient behaviour of $\sigma(0,t)$. This is also consistent with the observation that $\sigma$ remains constant during such a pressure pulse for larger $\rho_{C,b}$, since $\tau_{\rm reac} \propto \rho_{C,b}^{-1}$, such that the system is no longer reaction-limited for increased counter ion concentration. The  channel dimensions $H=1\mu$m and $L=40\mu$m are for computational reasons smaller than in the experiment, although the aspect ratio is the same. We checked that this time dependence is hardly dependent on $L$ and $H$ for fixed pressure drop amplitude $\Delta p=0.5~{\rm bar}$ and aspect ratio $L/H=40$ \cite{SIsize}. The similarity between the time-dependent experimental SFG-signal and $\sigma(x=0,t)/\sigma_{\rm eq}$ is striking, except perhaps for the strong short-time relaxation immediately after switching off the flow, which is present in our calculations (see Fig. \ref{fig:sc}) but absent in the experiment (inset). For comparison, Fig. \ref{fig:sc} also shows the surface charge for the case of a non-conducting Stern layer with $D_s=0$ (dotted red), which is virtually indistinguishable from $\sigma_{eq}$. By increasing the desorption rate, such that the system becomes less reaction-limited, the transient behaviour speeds up and the steady-state approaches the equilibrium state, as can be observed from the dashed line in Fig. (\ref{fig:sc}) \cite{SIrates}.

In conclusion, we apply the classical PNPNS equations (\ref{eq:equations}) to pressure-driven flow through a channel with newly formulated boundary conditions for out-of-equilibrium charge regulation and a conducting Stern-layer. For realistic system parameters, in particular for silica surfaces, this theory predicts a strong flow-induced heterogeneity of the surface charge and the $\zeta$-potential, even for a chemically homogeneous silica surface with Du$\ll 1$. The traditional Helmholtz-Smoluchowski relation (\ref{eq:smsurf}) for the streaming potential, which assumes a laterally constant $\zeta$-potential, breaks down for these regulating and conducting surfaces, provided the system is reaction-limited, i.e. the chemical reaction is the slowest process. In this reaction-limited regime, a nonzero conductive flux in the Stern layer must be largely compensated by an opposite diffusive surface flux (i.e. by a heterogenoeus surface charge) in order to prevent steady-state charge accumulation at the edges due to slow reaction kinetics. The resulting surface charge profile has a reduced lateral average $\langle\sigma\rangle$ compared to equilibirum. Our theory also provides a microscopic picture for measurements on the full time-dependence of the the relaxation dynamics of the surface charge after switching on and off a tangential flow \cite{lis}.  We have therefore shown that the combination of a non-zero surface conduction and (s)low chemical ad- and desorption rates can have dramatic impact on the interpretation of electrokinetics in micro- and nano-fluidic experiments, for which the surface charge and $\zeta$-potential are a vital component. We expect that these or similar mechanisms also play a role in electro-osmotic and diffusio-osmotic phenomena, which are interesting topics for future research in the context of e.g. blue energy harvesting \cite{blauw1,blauw2} and catalysis \cite{lammertink}.

\begin{acknowledgments}
We acknowledge fruitful discussions with F. Mugele and I. Siratanu, and thank M. Bonn who kindly provided us with the experimental data of Fig. 2D of Ref. \cite{lis}. This work is part of the D-ITP consortium, a program of the Netherlands Organisation for Scientific Research (NWO) that is funded by the Dutch Ministry of Education, Culture and Science (OCW). This work is a part of an Industrial Partnership Program of the Netherlands Organization for Scientific Research (NWO) through FOM Concept agreement FOM-15-0521. Financial support was provided through the Exploratory Research (ExploRe) programme of BP plc. J. C. E. acknowledges financial support from the European Union's Horizon 2020 programme under the Marie Skłodowska-Curie grant agreement No. 795377. 
\end{acknowledgments}

\bibliographystyle{apsrev4-1}


%

\newpage

\onecolumngrid
\section{Supplementary Materials to 'Flow-induced surface charge heterogeneity in electrokinetics due to Stern-layer conductance coupled to reaction kinetics'}

\twocolumngrid

\section{Dynamical density functional theory}
We consider a system consisting of bulk (region $\mathcal{R})$ and a surface (given by a region $\mathcal{S}$, with bulk ions (densities $\rho_a({\bf r\,})$, with ${\bf r} \in \mathcal{R}$) and surface ions (surface densities $\sigma({\bf r_s})$, with ${\bf r_s} \in \mathcal{S}$) that are bound to $\mathcal{S}$ via a chemical reaction. Here $i$ labels the bulk ion species. We will derive the equations using Density Functional Theory (DFT) \cite{DFTBook,DFTEvens}. It is convenient to split the intrinsic Helmholtz free energy functional $\mathcal{F}[\{\rho_i\},\sigma]=\mathcal{F}_b[\{\rho_i\}]+\mathcal{F}_s[\sigma]$, in a bulk contribution (first term) and a surface contribution (second term). Define the bulk charge density as $e\rho_e({\bf r})$ with $\rho_e({\bf r})=\sum_i z_{i}\rho_i({\bf r})$. The symbols for constants and material properties used are the same as in the Letter. Within mean field, the bulk functional is given by,
\begin{equation}
\begin{aligned}
\beta\mathcal{F}_b[\{\rho_i\}]=&\sum_i\int_\mathcal{R} \dif^3 {\bf r }\rho_i({\bf r})[\log\left(\rho_i({\bf r})\Lambda_i^3\right)-1]+\\
&\frac{1}{2}\int_\mathcal{R} \dif^3 {\bf r}\, \rho_e({\bf r\,})\phi({\bf r}), 
\end{aligned}
\end{equation}
with $\phi({\bf r},t)=\beta e\psi({\bf r},t)$ the dimensionless electrostatic potential. We describe the surface contribution as a two-dimensional lattice gas,
\begin{equation}
\begin{aligned}
\beta\mathcal{F}_s[\sigma]=&\sum\int_{\mathcal{S}} \dif^2{\bf r}_s \Big{[} {}\sigma({\bf r_s})\log\left(\sigma({\bf r_s})\Gamma^{-1}\right)\\&+
\left(\Gamma-\sigma({\bf r}_s)\right)\log\left((\Gamma-\sigma({\bf r}_s))\Gamma^{-1}\right)\Big{]} \\
&+\frac{1}{2}z_{\sigma}\int_{\mathcal{S}} \dif^2{\bf r}_s \ \sigma({\bf r}_s)\phi({\bf r}_s).
\end{aligned} \label{eq:surffunc}
\end{equation}
Here, $\Gamma$ is the surface density of chargeable sites and $z_{\sigma}$ the valency of the surface charges.  There is no free energy of binding included in Eq. (\ref{eq:surffunc}), since we are interested in out-of-equilibrium processes. We include this in the continuity equation to be given later. The type of chemical reactions that we are interested in, is the chemisorption of ions on a chargeable surface. As a model sample we look at the charging of a single neutral site SC by the desorption of an cationic counter ion C$^+$, given by SC $\leftrightharpoons$ S$^+$+C$^+$. Such a charging process is described by the rate equation (for convenience we will assume that the chemical reaction consists of a single elementary step),
\begin{equation}
\frac{\dif \{\mathrm{S}\mathrm{C}\}}{\dif t}=k^\text{des}\{\mathrm{S}\mathrm{C}\}-k^\mathrm{ads}[\mathrm{C}^+]\{\mathrm{S}^-\}. \label{eq:rate}
\end{equation}
Here the curly brackets indicate a surface density, and a square bracket a bulk density. The adsorption rate is denoted by $k^\text{ads}$, while $k^\text{des}$ denotes the desorption rate. We can in principle estimate these quantities using Arrhenius' law. More complicated rate equations can be investigated if the precise reaction mechanism is known. In equilibrium the LHS of Eq. \eqref{eq:rate} is zero, and we retrieve the equilibrium constant $K_{C}\equiv\{\mathrm{S}^-\}[\mathrm{C}^+]/\{\mathrm{SC}\}=k^\text{des}/k^\text{ads}$. The continuity equation in bulk is given by
\begin{equation}
\frac{\dif\rho_i({\bf r},t)}{\dif t}=-\nabla\cdot{\bf J}_i({\bf r\,},t),\quad{\bf r}\in\mathcal{R}\label{eq:contbulk},
\end{equation}
with ${\bf J}_i$ the bulk flux of ion species $i$. Note that we used the full (material) derivative of $\rho_i({\bf r},t)$ instead of the partial in order to account for advection. 
The bulk current ${\bf J}_{i}({\bf r},t)$ can be derived  using Dynamical DFT \cite{DDFT},
\begin{equation}
\begin{aligned}
{\bf J}_{i}({\bf r},t)&=-D_{b,i}\rho_i({\bf r},t)\nabla\left(\left. \frac{\delta\beta\mathcal{F}_b\left[\rho_{i}\right]}{\delta\rho_{i}({\bf r})}\right\vert_{\substack{\rho_{i}({\bf r},t)}}\right)\\
&=-D_{b,i}\left(\nabla\rho_i({\bf r},t)+z_i\rho_i({\bf r},t)\nabla\phi({\bf r},t)\right), 
\end{aligned}\label{eq:bulkcurrent}
\end{equation}
where we introduced diffusion coefficients $D_{b,i}$ for the ions in water.
For $\sigma({\bf r}_s)$ on the surface $\mathcal{S}$ the continuity equation reads
\begin{equation}
\frac{\partial\sigma({\bf r}_s,t)}{\partial t}=-\nabla_\mathcal{S}\cdot{\bf j}_{\sigma}({\bf r}_s,t)+ R({\bf r}_s,t),\quad{\bf r}_s\in\mathcal{S}. \label{eq:contsurf}
\end{equation}
where $R$ is the production rate of surface charges and ${\bf j}_{\sigma}$ the (2D) flux of surface charges. We have implemented a type B dynamic model because the total number of ions (on surface plus in the water) is conserved. Note that the divergence in Eq. \eqref{eq:contsurf} is a two-dimensional divergence. For example, for a flat plate in the $xy$ plane, we have $\nabla_\mathcal{S}=(\partial_x,\partial_y)$. 
From the example of Eq. \eqref{eq:rate} we can infer,
\begin{equation}
R({\bf r}_s,t)=-k^\text{ads}\left(\Gamma-\sigma({\bf r}_s,t)\right)+k^\text{des}\rho_{\rm C}({\bf r}_s,t)\sigma({\bf r}_s,t).
\label{eq:rateeq}
\end{equation}
Analogously to the bulk equation, the surface current is given by,
\begin{equation}
\begin{aligned}
{\bf j}_{\sigma}({\bf r}_s,t)&=-D_{s}\sigma({\bf r}_s,t)\nabla_\mathcal{S}\left(\left. \frac{\delta\beta\mathcal{F}_s\left[\sigma\right]}{\delta\sigma({\bf r}_s)}\right\vert_{\substack{\sigma({\bf r}_s,t)}}\right)\\
&=-D_s\left(\dfrac{\Gamma\nabla_\mathcal{S}\sigma({\bf r}_s,t)}{\Gamma-\sigma({\bf r}_s,t)}+z_{\sigma}\sigma({\bf r}_s,t)\nabla_\mathcal{S}\phi({\bf r}_s,t)\right), 
\end{aligned}\label{eq:surfcurrent}
\end{equation}
where we introduced diffusion coefficient $D_s$ for the ions adsorbed in the Stern layer. Furthermore, similarly for Eq. (\ref{eq:bulkcurrent}), we implicitly used the Einstein-Smolochowski relation. Finally, we close the above set of equations, by the Poisson equation (the potential field is generated instantaneously, no retardation):
\begin{equation}
\nabla^2\phi({\bf r},t)=-4\pi\lambda_B[\rho_e({\bf r},t)+z_{\sigma} \sigma({\bf r},t)f({\bf r})],
\label{eq:poisson}
\end{equation}
where $\lambda_B=\frac{\beta e^2}{4\pi\epsilon}$ is the Bjerrum length and the function $f({\bf r})$ with dimension inverse length encodes the location of the chargeable surface. For example, for a chargeable plate at $z=0$ it is given by $f({\bf r})=\delta(z)$. 

Lastly, to describe the fluid flow we employ the Navier Stokes equation with the incompressibility condition:
\begin{align}
&m\frac{\partial {\bf u}({\bf r},t)}{\partial t}+m({\bf u}({\bf r},t)\cdot\nabla){\bf u}({\bf r},t)=
-\nabla p({\bf r},t)\\ & +\eta\nabla^2 {\bf u}({\bf r},t)\nonumber
 +e\rho_e({\bf r},t){\bf E}({\bf r},t), \nonumber\\
&\nabla\cdot{\bf u}({\bf r},t)= \, 0. \label{eq:navstok}
\end{align}
Here ${\bf u}({\bf r},t)$ describes the fluid velocity field, $p(\bf{r},t)$ the pressure and ${\bf E}({\bf r},t)=-\nabla \psi({\bf r},t)$ the electric field. The last two terms represent the electric body force on the fluid due to migrating ions. 
Combining Eqs. (\ref{eq:contbulk}),(\ref{eq:rateeq}),(\ref{eq:contsurf}),(\ref{eq:bulkcurrent}),(\ref{eq:surfcurrent}),(\ref{eq:poisson}) \& (\ref{eq:navstok}) gives the set of governing equations. 

To couple the bulk and the surface we use a Robin boundary condition, which states that the production rate of surface charges is equal to the counter ion flux normal to the surface. Furthermore, we use the standard electric boundary condition as well as the no-slip boundary condition for ${\bf u}$. 
\begin{equation}
\begin{aligned}
{\bf n}_s\cdot{\bf J}_{\rm C}({\bf r}_s,t)&=-R({\bf r}_s,t),\quad{\bf r}_s\in\mathcal{S},\\
{\bf n}_s\cdot \nabla \phi({\bf r}_s,t)&=-4\pi\lambda_{\rm B} \sigma({\bf r}_s,t)\\
{\bf u}({\bf r}_s,t)&=0
\end{aligned}\label{eq:coupling}
\end{equation}
with $\bf n$ an inward pointing normal vector (into the fluid) and ${\bf J_{\rm C}}$ the counter ion flux. Lastly, we take all non-charged surfaces impermeable for all ions, and the charged surface impermeable for all ions expect the counter ion.

In equilibrium ($t\rightarrow\infty$), the net flux between bulk and surface vanishes, and hence by Eq. \eqref{eq:coupling}, we find that $R({\bf r}_s,t\rightarrow\infty)=0$. For our model of a single cation desorbing from a neutral surface, we retrieve the Langmuir adsorption isotherm,
\begin{equation}
\sigma({\bf r_s},t\rightarrow\infty)=\Gamma\left[1+\frac{\rho_C({\bf r_s},t\rightarrow\infty)}{K_{C}}\right]^{-1},\quad{\bf r_s}\in\mathcal{S}. \label{eq:langmuir}
\end{equation}
We can find the bulk counter ion density $\rho_C({\bf r})$ by the condition ${\bf J}_C({\bf r},t\rightarrow\infty)=0$, which from Eq. \eqref{eq:bulkcurrent} translates to 
\begin{equation}
\frac{\delta\beta\mathcal{F}_b}{\delta\rho_{C}({\bf r})}=\text{constant}=\mu_{\mathcal{R}},
\end{equation}
with $\mu_{\mathcal{R}}$ the chemical potential of the system, which in the grand canonical ensemble is given by the bulk ion and water reservoir. Strictly speaking, the above constant should include the external potential. The only external potential in our system is the hard wall potential, which we already have included via the integration limits (bounds of $\mathcal{R}$). The above equation implies that
\begin{equation}
\rho_i({\bf r},t\rightarrow\infty)=A_i\exp[-z_i\phi({\bf r},t\rightarrow\infty)]. \label{eq:boltzmann}
\end{equation}
For the grand canonical system $A_i$ is determined from $\mu_{\mathcal{R}}$, and can be set equal to the bulk salinity.

Note that Eqs. \eqref{eq:langmuir} and \eqref{eq:boltzmann} are internally consistent with the condition
\begin{equation}
\frac{\delta\beta\mathcal{F}_s}{\delta\sigma({\bf r}_s)}=\text{constant}.
\end{equation}
Alternatively, we could also have enforced ${\bf j}_{\sigma}({\bf r}_s)=0$, which gives a constraint on the surface charge,
\begin{equation}
\sigma({\bf r}_s,t\rightarrow\infty)=\Gamma\left[1+C_{\sigma} \, e^{z_{\sigma}\phi({\bf r}_s,t\rightarrow\infty)}\right]^{-1}, \quad{\bf r}_s\in\mathcal{S},
\end{equation}
for some constant $C_{\sigma}$. Therefore, both setting the source and flux for surface charges to zero gives the familiar Langmuir type adsorption isotherm used in (equilibrium) charge regulation schemes. Note here that if we set any 2 of ${\bf J}_i$, ${\bf j}_{\sigma}$ or $R$ to zero, that immediately implies that the third is zero also. This shows the internal consistency of the proposed theory. 

\section{Dimensionless form}

In order to numerically solve the equations, it is best to convert the governing equations to a dimensionless form. We scale all densities by the reservoir salt concentration $\rho_b$, $\rho_i({\bf r},t)=\rho_b n_i({\bf r},t)$ the distances by the Debye length $\lambda_D\equiv\left(\frac{2e^2\rho_b}{\epsilon k_{\rm B}T}\right)^{-1/2}$, ${\bf r}=\lambda_D {\bf r}'$, the velocity by a typical velocity $u_0$, ${\bf u}=u_0 {\bf v}$, the potential by the thermal voltage $k_{\mathrm{B}}Te^{-1}$, $\psi({\bf r},t)=k_{\mathrm{B}}Te^{-1}\phi({\bf r},t)$ and the surface charge by the site density $\Gamma$, $\sigma({\bf r}_s,t)=\Gamma s({\bf r}_s,t)$.  We scale time by the time it takes for an ion to diffuse over this Debye-length, $\tau_{dif}\equiv \lambda_D^2D_b^{-1}$, $t=\tau_{dif}t'$, with $D_b$ the diffusion constant of the dissolved ions which we take to be equal for all ions ($D_{b,i}=D_b$). This scaling provide the following set of differential equations for the quantities to be solved in this system,
\begin{equation}\label{eq:equations}
\begin{aligned}
\nabla'^2&\phi=-\frac{1}{2}\left(n_+-n_-+n_C\right)=-\frac{1}{2}n_e,\\
\dfrac{\partial n_i}{\partial t'}&=\nabla'\left(\nabla'n_i+z_in_i\nabla'\phi \right)-\dfrac{\lambda_D}{H}\text{Pe}\, {\bf u}\cdot \nabla'n_i \\
\dfrac{\partial s}{\partial t'}&=W\nabla'_{\mathcal{S}}\left(\frac{\nabla'_{\mathcal{S}}s}{1-s}+s\nabla'_{\mathcal{S}}\phi_s\right)+\text{Da}_2\left((1-s)-K^{-1}r_{C,s} s\right),\\
\dfrac{\partial {\bf v}}{\partial t'}&=-\dfrac{\lambda_D}{H}\text{Pe}({\bf v}\cdot\nabla'){\bf v}-\nabla' p+\text{Sc}\nabla'^2{\bf v}+B n_e\nabla'\phi,\\
\nabla'&\cdot {\bf v}=0,\\
\end{aligned}
\end{equation}
with $n_e=\sum_iz_in_i$ the net charge density. The subscript 's' is a shorthand notation that the quantity is to be evaluated on the surface. The relevant length scale for the PNP equations is the Debye length, which can be concluded from the absence of dimensionless terms from the Poisson equation. For the NS equation, however, the relevant length scale is $H$. In our case of a flat charged plate and translational invariance in the $y$-direction, the derivative operators reduce to the simple form $\nabla=(\partial_x,\partial_z)$ and $\nabla_{\mathcal{S}}=\partial_x$. Furthermore, we have rescaled the pressure with the factor $\frac{m\lambda_D}{u_0D_b}$ such that $p$ is also dimensionless. We have furthermore introduced some dimensionless numbers, most of which with a pre-existing name:
\begin{enumerate}
\item $\text{Sc}=\frac{\eta}{mD_b}$, the Schmidt number, the ratio between momentum and mass diffusivity,
\item $\text{Pe}=\frac{u_0H}{D_b}$, the Peclet number, the ratio between advective and diffusive transport
\item $\text{Da}_2=\frac{k^{\text{des}}\lambda_D^2}{D_b}=k^{\rm des}\tau_{dif}$, a secondary Damk\"ohler number defined as the ratio between the chemical and diffusive time scale
\item $\text{K}=\frac{k^{\text{des}}}{\rho_sk^{\rm ads}}=\frac{K_{\rm C}}{\rho_b}$, with $K_{\rm C}$ the chemical equilibrium constant which determines the equilibrium surface charge
\item B$=\frac{k_{\mathrm{B}}T\rho_s\lambda_D}{mD_bu_0}$, the body force number, the ratio between the ionic body forces and the inertia of the fluid
\item $W=\frac{D_s}{D_b}$, the ratio between the bulk and surface diffusion constants.
\end{enumerate}

The Peclet number plays a role in transport equation (PNP in this case), and shows the significance of convection is determining the concentration profiles. Mathematically it represents the ratio between the local time derivative and the convection term. Analogously, the Schmidt number is the ratio between the viscous term and the local time derivative. The Damk\"ohler number is a measure that indicates if the system is diffusion or reaction limited. It is the ratio between the typical time it takes for a ion to adsorb and the time it takes for it to diffuse out of the double layer (and thus out of reach of the surface). The body force number $B$ signifies if the fluid flow is significantly influence by the movement of the ions. For the calculations we furthermore set $u_0=\frac{D}{\lambda_D}$ which automatically sets $\frac{\lambda_D}{H}$Pe=1. Although computationally convenient, we will not implement this for the coming analysis as there is no way of ensuring that $|{\bf v}|$ is of the order 1 in that case.

By estimating the value of the numbers Sc, Pe, K and B we can gain insight in the importance of certain terms. However, one must keep in mind that the terms associated with these numbers might not be of the order unity, if for example the derivatives are of a different scale. To simplify the calculations, we will analyse the Navier-Stokes equation. Many of the parameters are set by the properties of the bulk water, $\rho_b=1$ mM, $H=0.5~$ mm, $m=10^3$ kg m$^{-3}$, $D=10^{-9}$m$^2$/s, $T=293 K$ and $\eta=10^{-3}$Pa$\,$s. The effects described in the Letter emerge for $u_0\approx \, 0.01$m/s, which is one to two orders of magnitude lower than the fluid velocities reported by Lis et al. This can be attributed to the different system size, which is two orders of magnitude larger. These values result in Sc=10$^3$, $\frac{\lambda_D}{H}$Pe=$10^{-1}$ and B$\approx$2500. However, we must also implement that the fluid velocity varies over lengths of the order of $H$, while the potential $\phi$ varies over lengths of the order $L$. In order to fairly compare the different terms of the Navier-Stokes equation, we must compare the numbers $\frac{\lambda_D^2}{H^2}$Pe, $\frac{\lambda_D^2}{H^2}$Sc and B$\frac{\lambda_D}{L}$. Using typical values of $H=1\,\mu$m and $L=40\,\mu$m, we obtain $\frac{\lambda_D^2}{H^2}$Pe=10$^{-3}$, $\frac{\lambda_D^2}{H^2}$Sc=10 and $\frac{\lambda_D}{L}$B$\approx 0.5$. Thus when solving the Navier-Stokes equation we can safely ignore the inertial term and the ionic body forces compared to the viscous force. The resulting equation, the Stokes equation, gives rise to a Poisseulle flow. This allows the fluid velocity to be solved first, and to be used as input for the PNP equation. Note that although the experiments are with $H$ and $L$ typically two orders of magnitude higher, the same value of $H/L$ is used so the ratio between the different terms is unaffected.
\begin{figure}[!ht]
\centering
\includegraphics[width=0.5\textwidth]{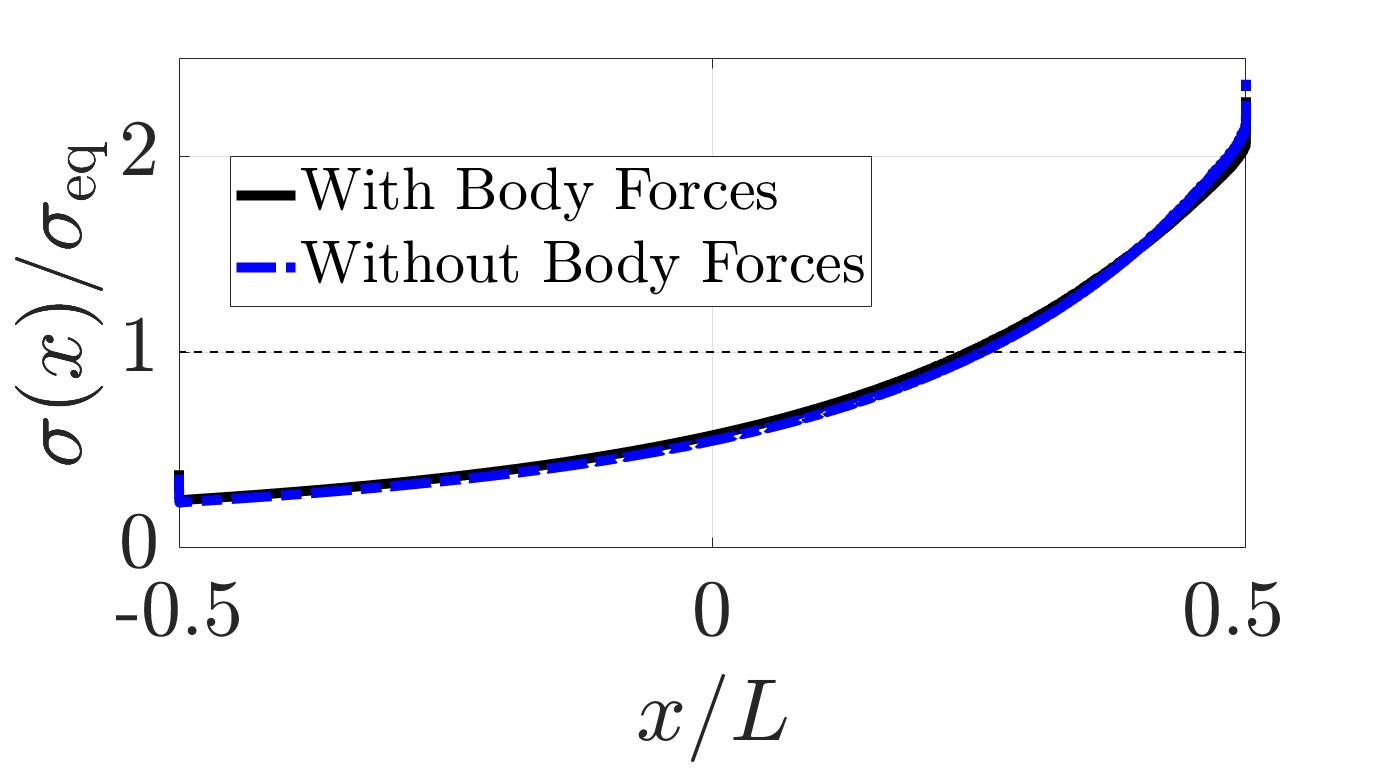}
\caption{The surface charge profile compared to equilibrium for a calculation including the body forces of Eq. (\ref{eq:equations}) (blue dashed line) and without the body forces (full black line). Both were calculated using the same parameters representing silica at pH=6.5, and using $\Delta p=0.5$~bar.}\label{fig:fullysolved}
\end{figure}
In order to confirm this prediction we solved the full set of equations. Fig. (\ref{fig:fullysolved}) shows the surface charge profile $\sigma(x)$ for both cases, one where the fluid flow is solved first and used as input for the other equations, and one where all equations are solved simultaneously. The advantage of the former is that the calculation times are a factor 5 shorter. The two cases are nearly indistinguishable, confirming the analyses above, so we can safely use the faster method.

\section{Simulation domain}

\begin{figure}[!ht]
\centering
\includegraphics[width=0.5\textwidth]{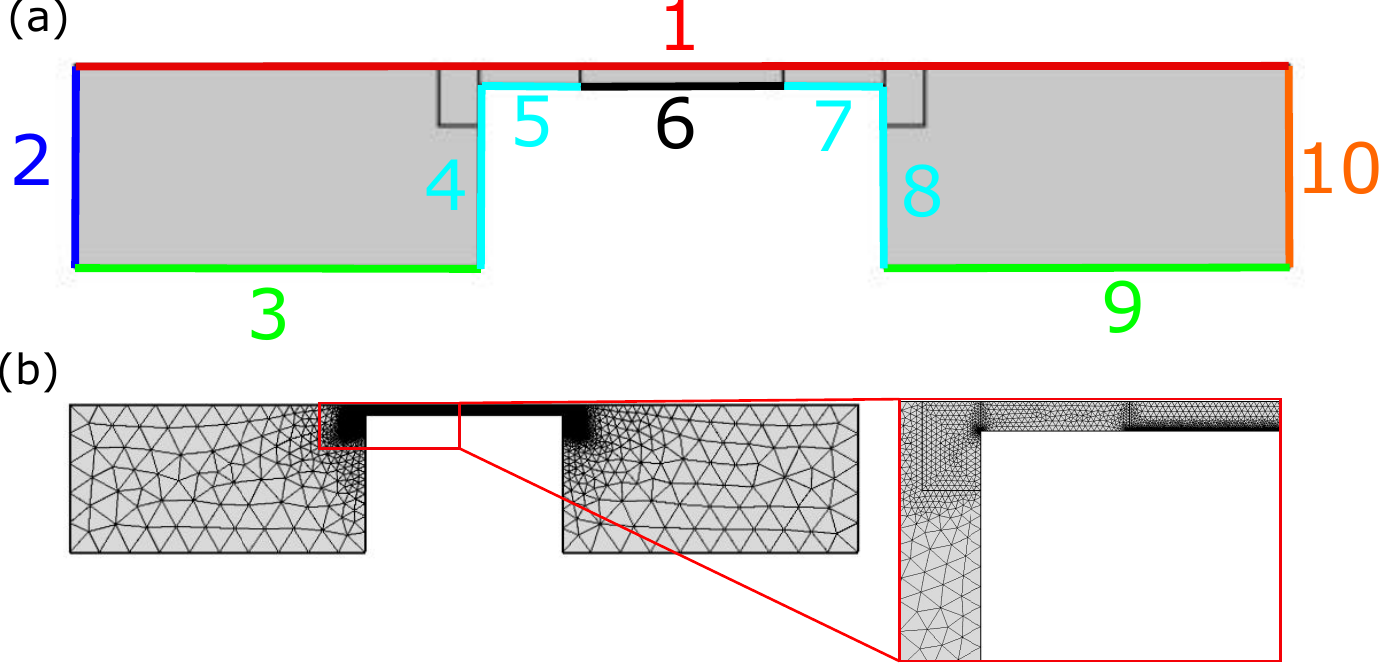}
\caption{Domain on which the governing equations are solved numerically: (a) domain with the boundaries marked (boundary conditions explained in the text and (b) a representation of the typically used mesh. The system size used is $H=1\,\mu$m and $L=5\,\mu$m. }\label{fig:domain}
\end{figure}

We solved the set of dimensionless equations \ref{eq:equations} on a geometry shown in Fig. \ref{fig:domain}. This domain is furthermore only two dimensional, since we have translation symmetry in the $y$ directoin. Since the system is symmetric in the plane at $z=\frac{1}{2}H$, only half of the system has to be included in the calculations. The boundary condition on this plane is then simply that all normal derivatives are zero.

For each simulation domain we can write down the boundary conditions.
\begin{enumerate}
\item[1] ({\color{red} Red}) All normal derivatives and velocities are zero, ${\bf n}\cdot \nabla \rho_i=0={\bf n}\cdot \nabla \psi={\bf n}\cdot {\bf u}$
\item[2] ({\color{blue} Dark blue}) Inlet reservoir, where we fix the pressure $p=\Delta p$, salinities $\rho_{\pm}=\rho_s$ \& $\rho_{C}=\rho_{C,b}$ and potential $\psi=0$
\item[3,9] ({\color{green} Green}) To simulate an infinite bulk, we impose slip boundary conditions $u_z=0$ on the side of the bulk, and impose a zero-flux boundary condition $\mathbf{z}\cdot \mathbf{J}_i=0$ for the ion concentrations.
\item[4,5,7,8] ({\color{cyan} Cyan}) Hard walls with no-slip and no-flux boundary conditions, $\mathbf{u}=0$ and $\mathbf{n}\cdot \mathbf{J}_i=0$
\item[6] ({\textbf Black}) The charged wall. The same boundary conditions for $\rho_{\pm}$, and ${\bf u}$ as 4/5/7/8, but for the counter ion we impose the chemical rate equation ${\bf n_s \cdot J}_C=-R$ with $R$ given by Eq. (\ref{eq:rateeq}), and for $\psi$ we impose the standard electrostatic boundary condition as in Eq.  (\ref{eq:coupling}).
\item[10] ({\color{orange} Orange}) Outlet reservoir. To simulate an infinite bulk we impose that all diffusive fluxes and the electric field are zero, ${\bf n}\cdot \nabla \rho_i=0={\bf n}\cdot \nabla \psi$, while we fix $p=0$.
\end{enumerate}

In order to solve for the surface charge density $\sigma$, we couple this two dimensional domain to a one dimensional domain where we solve the governing equations for $\sigma$, Eqs. (\ref{eq:surfcurrent},\ref{eq:contsurf}). This coupling can be achieved by the general (or linear) extrusion operator of COMSOL, which projects the value of $\rho_{C,s}$ from the 2D geometry on the 1D geometry, and the value of $\sigma$ from the 1D geometry on the 2D geometry. The additional boxes created at the entrance and exit of the channel help to refine the mesh in these regions, where a finer mesh is needed than in the in- and outlet reservoirs (see Fig. \ref{fig:domain}(b)). At the former, mesh elements should be no larger than 0.1$H$, and at the corner we chose a slightly refined mesh to reduce numerical inaccuracies. Furthermore, at boundary 6 a very fine mesh is required as the full Electric Double Layer must be resolved (since $\rho_{C,s}$ must be determined with accuracy at the surface). In the double layer, a grid of no lower than 8 points per Debye length was required for consistently reliable results, although the specific grid might vary with system size. This mesh smoothly transitions to the coarser mesh away from the charged surface towards the center of the channel. For typical values of $H=1\mu$m and $L=30\mu$m, this results in a number of gridpoints of the order of 10$^5$ mesh points and a calculations time of the order of an hour for the full dynamical calculations, i.e. equilibration, transition to steady state upon application of pressure pulse and re-equilibration after the pressure drop has subsided to zero (calculation time depends strongly on value of $L$ and $H$).

\section{The system size}

Experimental system sizes of streaming potential or streaming current set-ups can vary largely in size, but commonly occur on larger scales than the sub-millimetre sized system consider in this Letter. For example, in the experiment of Lis et al., $H=0.5~$mm and $L=20~$mm.
\begin{figure}[!hb]
\centering
\includegraphics[width=0.5\textwidth]{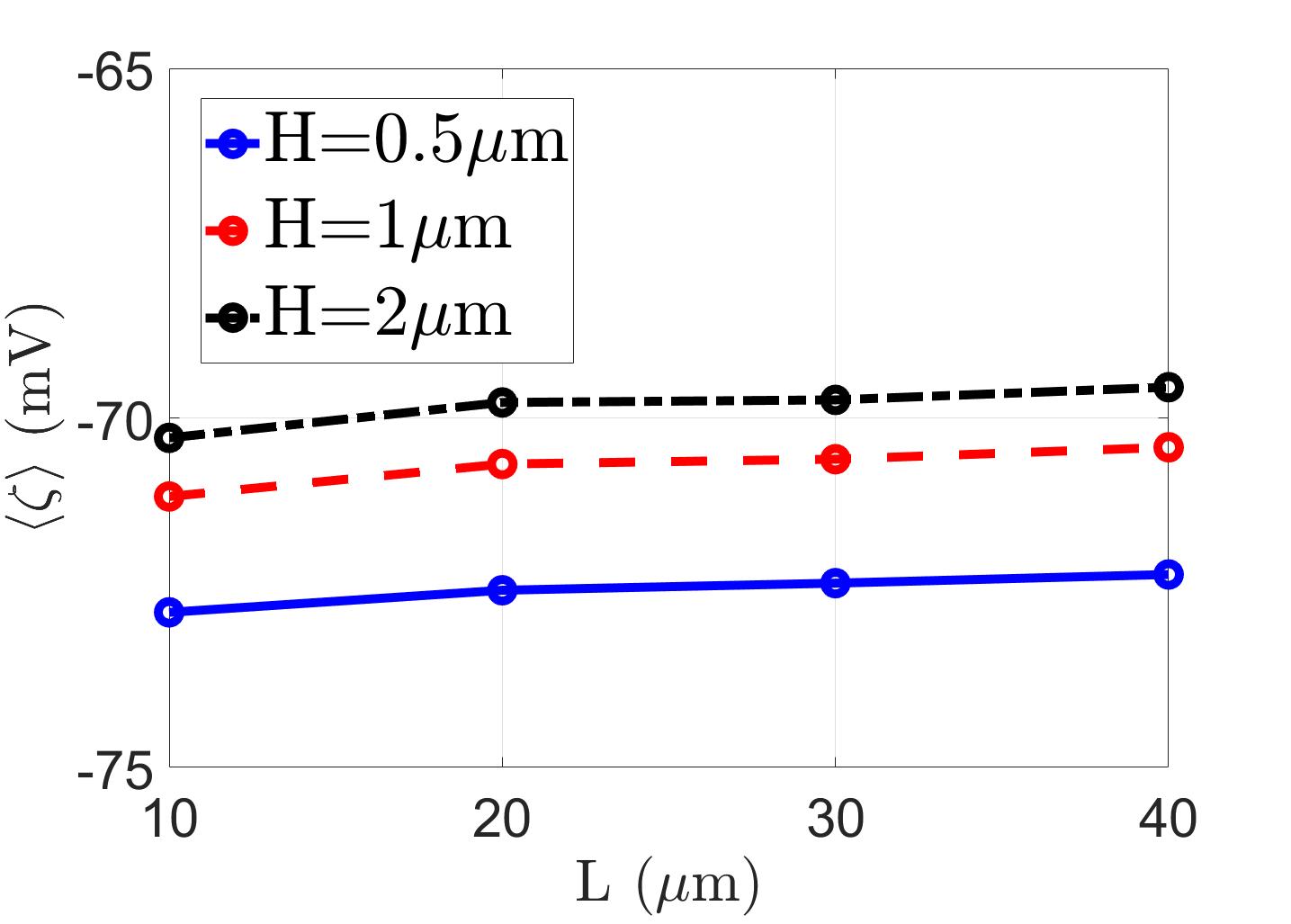}
\caption{The average $\zeta$-potential $\langle\zeta \rangle$ of the charged channel surface in the steady state as a function of the channel length $L$ and channel height$H$. Parameters were chosen here too in order to represent silica at pH=6.5.}\label{fig:SystemSize}
\end{figure}
\begin{figure}[!ht]
\centering
\includegraphics[width=0.50\textwidth]{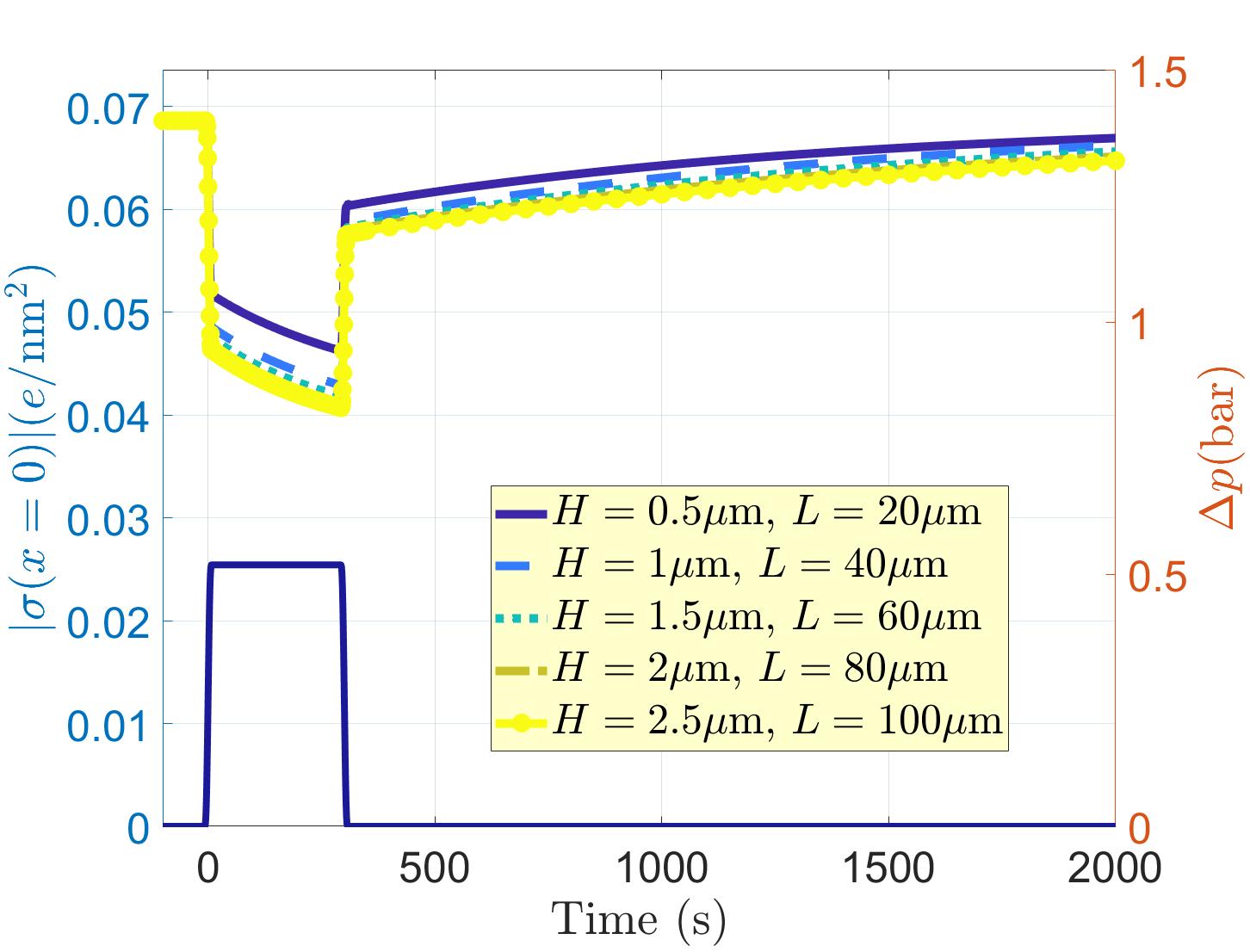}
\caption{The surface charge in the middle of the surface, $\sigma(x=0,t)$, as a function of time for $L=20,40,60,80,100~\mu$m while maintaining $L/H=40$. Parameters were chosen here too in order to represent silica at pH=6.5.}\label{fig:LisSystemSize}
\end{figure}
These sizes are numerically out of reach, however, since the smallest length scale of our system is the Debye length $\lambda_D\simeq 10$nm. Therefore we varied $H$ and $L$ over values that are numerically more feasible: $H=\{0.5,1,2\}\mu$m and $L=\{10,20,30,40\}\mu$m with a fixed $\Delta p$. 

Fig \ref{fig:SystemSize} shows the average $\zeta$-potential $\langle\zeta \rangle$ for these values of $H$ and $L$. Although $\langle\zeta \rangle$ does depend slightly on both $L$ and $H$, the variation is only minimal. This therefore suggests that the observed heterogeneity persists even for larger system sizes. The dependency of $H$ can be further understood via the Duhkin number Du, which is inversely proportional to $H$. For larger $H$, the relative contribution of the increased conductivity of the surface decreases, which in turn effects the streaming potential and thus $\langle\zeta \rangle$.

Similarly, we can inspect the effect of system size for the transient behaviour. In Fig. \ref{fig:LisSystemSize} shows the transient behaviour for several values of $L$ and $H$. The ratio $L/H=40$ is held fixed, as it is the same ratio as in the experiment of Lis et al. Fig. \ref{fig:LisSystemSize} shows the effect of scaling up the system. While there is some effect for smaller system sizes, the transient behaviour seems to approach an asymptote as $L$ increases (the difference between $L=80~\mu$m and $L=100~\mu$m is hardly discernible). This suggests that, when $\Delta p$ is held fixed, scaling up the system should not significantly impact the properties of both the transient and steady state behaviour, and that the results presented here are also valid for larger systems.

\section{Smoluchowski Equation}

\begin{figure}
\centering
\includegraphics[width=0.5\textwidth]{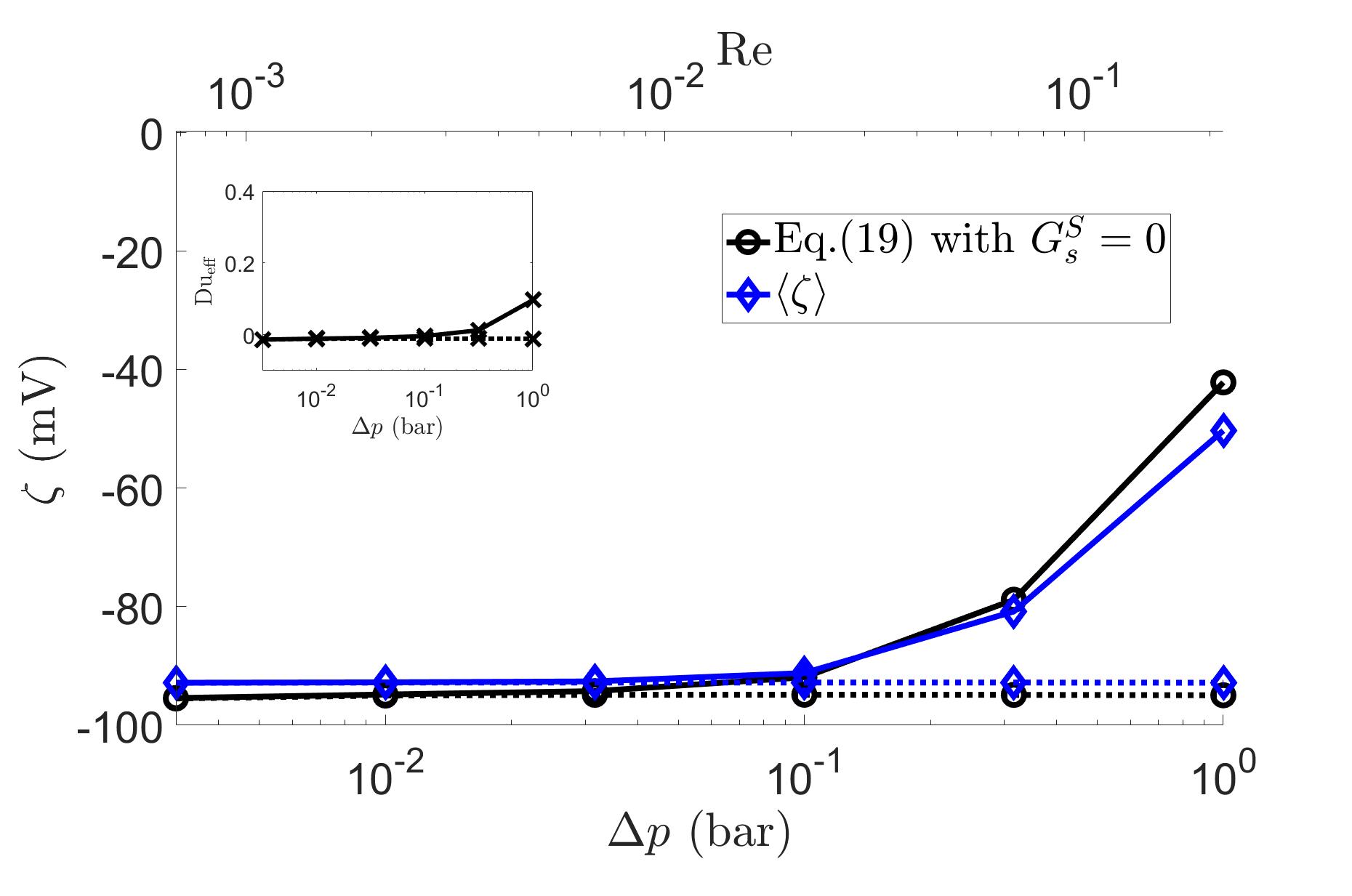}
\caption{The $\zeta$-potential as a function of the pressure drop $\Delta p$ and Reynolds number Re, as predicted by Eq. (\ref{eq:smsurf}) (black circles), with $G_s=G_S^d$, and the numerically calculated average value, $\langle\zeta\rangle$  (blue diamonds). The full and dotted lines represent calculations with $D_s=D$ and $D_s=0$ respectively. The difference at low $\Delta p$ is due to the diffuse layer conduction, $G_{s,d}$. The inset shows the Duhkin number as calculated by the ratio of the calculated and predicted $\langle\zeta\rangle$. Parameters were chosen to represent silica at pH=6.5, with an equilibrium $\zeta$-potential $\zeta_{\rm eq}\approx -93$ mV.}\label{fig:linres}
\end{figure}

The derivation of the Smoluchowski equation, 
\begin{equation}\label{eq:smsurf}
\Delta \Phi_S=\dfrac{-\zeta \epsilon}{\eta G}\Delta p,
\end{equation}
assumes a spatial homogeneity of both $E_x$ and $\zeta$. Therefore, in light of the discussed results, we can no longer simply assume the applicability of this equation to our system. 
In order to test the applicability, we calculated $\Delta\Phi_S$ at different $\Delta p$, and determined the value of $\zeta$ as predicted by Eq. (\ref{eq:smsurf}). Fig. \ref{fig:linres} shows this prediction in the case of $D_s=D$ (black full line, circles) and the case of $D_s=0$ (black dotted line, circles). The parameters were chosen to represent silica at pH=6.5 (see above), using a system size of $L=30 \, \mu$m and $H=1\,\mu$m. We compare this to the calculated average $\zeta$-potential, $\langle \zeta \rangle$, for both $D_s=D$ (blue full line, diamonds) and $D_s=0$ (blue dotted line, diamonds). For the prediction using Eq. (\ref{eq:smsurf}) we set $G_s=G_s^d$, where the diffuse layer conductivity $G_s^d$ can be calculated using Bikerman's expression \cite{Delgado}. The data is clearly independent of $D_s$ for $\Delta p\rightarrow 0$, as the difference between the prediction and the calculated value is small. Eq. (\ref{eq:smsurf}) seems to slightly overpredict the $\zeta$-potential, but the effect is small (a few miliVolts) showing that Eq. (\ref{eq:smsurf}) is a good approximation. This, however, also shows that for small $\Delta p$ the conductivity via the Stern layer is small and that (in the chosen parameter regime) the electric current via the Stern layer is negligible. We can explain this independence on $D_s$ by considering the strong asymmetry in time-scales, and thus that the charge exchange between surface and water is too slow to contribute significantly to the total charge current. For higher $\Delta p$, we see that $\langle \zeta\rangle$ starts to deviate from its equilibrium value if $D_s\neq 0$, while for $D_s=0$ we have that $\langle \zeta\rangle=\zeta_{eq}$ and Eq. (\ref{eq:smsurf}) remains valid. Nonetheless, Eq. (\ref{eq:smsurf}) remain to give an accurate prediction, but now of $\langle \zeta\rangle$ and \emph{not} $\zeta_{eq}$. The prediction becomes less precise as $\Delta p$ grows, which can be converted to an effective contribution of the Stern layer to the Duhkin number,
\begin{equation}
1+2{\rm Du}^S_{eff}=\dfrac{-\Delta \Phi_S \eta G_b}{\epsilon \Delta p}; \qquad G_d=G_b+G_s^d/H
\end{equation} 
The obtained Du$^S_{eff}$, shown in the inset, at low $\Delta p$ is negligible. The inset furthermore shows that the effective Duhkin number increases as $\Delta p$ increases. This, however, does not necessarily indicate an increase in surface conduction, but more likely is caused by the decreasing validity of Eq. (\ref{eq:smsurf}), which is not equipped to deal with lateral heterogeneities. However, contrary to previous surface conduction studies, where Du is the main focus, Du does not play a key role here. The observed effects are maintained even at higher $H$: increasing $H$ but keeping $L/H$ and $\Delta p$ fixed (which decreases Du) does not alter the results significantly (see Fig. \ref{fig:SystemSize} \& \ref{fig:LisSystemSize}), further signifying the insignificant role player by Du for the presented results.

Fig. \ref{fig:linres} clearly shows Eq. (\ref{eq:smsurf}) no longer adequately predicts the equilibrium $\zeta$-potential at high $\Delta p$, even if corrected for the extra conductivity via the surface (i.e. by adjusting $G_s$). Instead, Eq. (\ref{eq:smsurf}) gives a good approximation of $\langle\zeta\rangle$, the average value of the $\zeta$-potential, which in the current discussion is a function of the pressure drop. This is to be expected, since the streaming potential is the integrated effect of the advection in the EDL, which is in turn determined by the \emph{local} $\zeta$. Note that also Du$_{eff}^S$ is a function of $\Delta p$, as shown in the inset. What is surprising is that $\langle\zeta\rangle <\zeta_{eq}$ (this also implies that $\langle\sigma\rangle<\sigma_{eq}$), and thus that the surface has a smaller net charge in the steady state than in equilibrium. The large $E_x$ creates a large excess of surface charges on one side of the channel, where the increased adsorption rate of counter ions ($k^{\rm ads}\sigma\rho_{C,s}$) is larger than the increased desorption rate ($k^{\rm des}(\Gamma-\sigma)$). To exclude hydrodynamic effects as a cause we also plotted the correspond Reynolds number value, Re=$\frac{m u_{\rm max} H}{\eta}=\frac{mH^3\Delta p}{16\eta^2L}$, where we have used that the maximum fluid velocity of a Poiseuille flow is given by $u_{\rm max}=\frac{H^2\Delta p}{16\eta L}$. Under all circumstances Re $<1$, so we can expect a fully developed Poiseuille flow along the charged surface.

\section{The chemical rates}

\begin{figure}[!ht]
\centering
\includegraphics[width=0.5\textwidth]{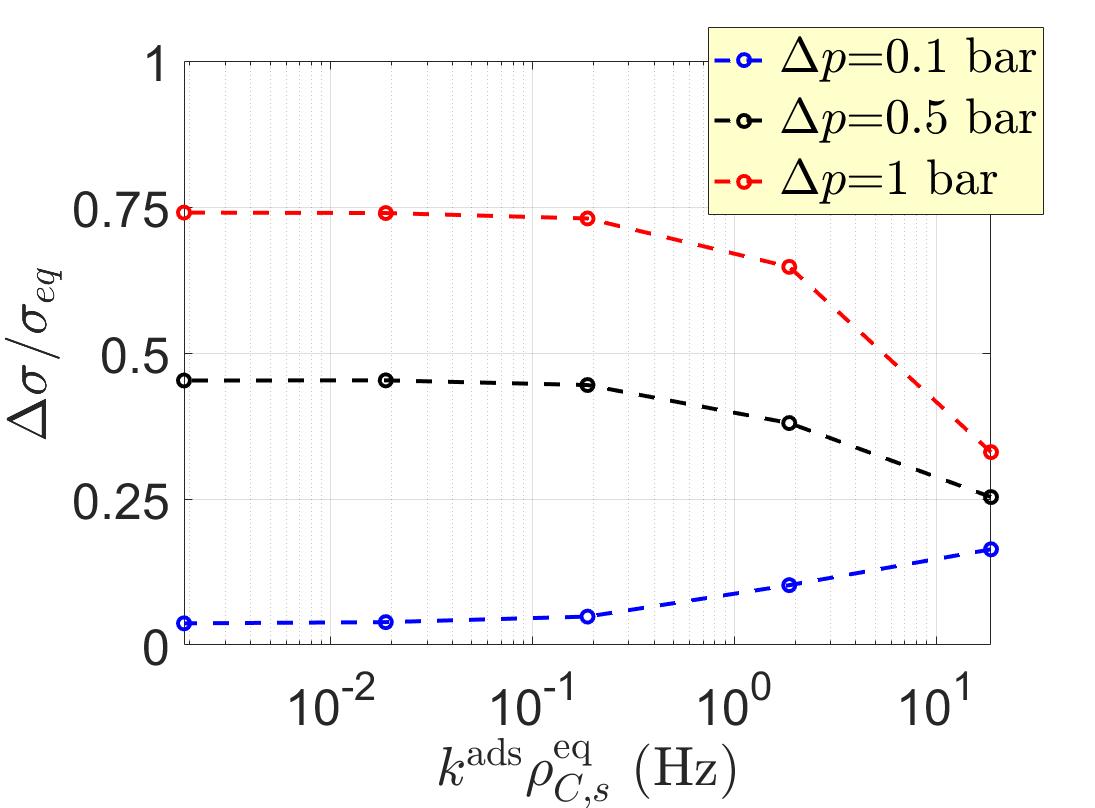}
\caption{Difference between steady state and equilibrium surface charge at the center of the channel as a function of the chemical desorption rate $k^{\rm des}$, for $\Delta p=0.1$ (blue), 0.5 (black) and 1 (red) bar.}\label{fig:Deltas}
\end{figure}

As argued in the Letter, the observed heterogeneous surface charge profiles relies on the balance of the timescales of the system, and more specifically that the chemical timescale $\tau_{\rm reac}$ is the largest. An order of magnitude of $\tau_{\rm reac}$ can be obtained from the continuity equation for $\sigma$ Eq. (\ref{eq:contsurf}). Assuming a laterally constant $\rho_{C,s}$ and $\sigma$ and $j_{\sigma}=0$, the continuity equation of $\sigma$ reverts to a simple linear differential equation for $\sigma(t)$. The governing timescale is then straightforwardly deduced as $\tau_{\rm reac}= k^{\rm des}\left(1+\rho_{C,s}/K\right)\simeq k^{\rm ads}\rho_{C,s}$. Here we have approximated $\rho_{C,s}$ with its equilibrium value, $\rho_{C,s}^{\rm eq}=\rho_{C,b} e^{-e \zeta_{\rm eq}/k_{\rm B}T}$ such that $\rho_{C,s}/K\simeq\rho_{C,b} e^{-e \zeta_{\rm eq}/k_{\rm B}T}/K \gg 1$ since $\rho_{C,b}/K\simeq 1$ ($\zeta_{\rm eq}=-93$~mV).  To investigate the effect of the chemical reaction rates $k^{\rm ads}$ and $k^{\rm des}$ , we repeated the calculation of Fig. (2) of the Letter but at different $k^{\rm ads}$. Higher rates than the largest value shown (10 Hz)  exhibited numerical convergence problem). Fig. \ref{fig:Deltas} shows the difference between the steady state and equilibrium value of the surface charge in the middle of the surface, $\Delta \sigma \equiv \sigma(0)|_{\rm steady state}-\sigma_{\rm eq}$, for $\Delta p =0.1, 0.5$ and 1~bar.
Fig. \ref{fig:Deltas} shows that, as a function of the desorption rate, the systems seems to depart from the reaction-limited regime. The value for $\tau_{\rm reac}$ for which this transition occurs is independent of $\Delta p$. Since $j_{\sigma}$ and thus $\tau_{\rm cond}$ depends directly on $\Delta p$, this is consistent with our timescale analysis. For very small rates, $\Delta \sigma$ is fixed by $\Delta p$, or more specifically by the generated electric field and surface current $j_{\sigma}$. In the limit $k^{\rm ads}\rho_{C,s}\rightarrow 0$ the source term ${\bf n}\cdot {\bf J}_e$ vanishes, so in a steady state the surface flux must also vanish (a constant $j_{\sigma}$ is prohibited as the charged surface is finite) and the generated profile $\sigma(x)$ is such that $j_{\rm cond}=D_s\frac{e\sigma}{k_{\rm B}T}\frac{\partial \psi_s}{\partial x}$, the conductive surface flux, and the diffusive surface flux, $j_{\rm dif}=-D_s\frac{1}{1-\sigma/\Gamma}\frac{\partial\sigma}{\partial x}$, balance each other. For faster rates, however, this is possible, so $j_{\rm cond}$ and $j_{\rm dif}$  don't necessarily have to balance each other, the difference being equal to the flux into the water (${\bf n}\cdot {\bf J}_C$). In this regime, the chemical reaction is fast enough such that the surface charge does not change from its equilibrium value. In terms of the timescales discussed in the Letter, we see that the chemical timescale $\tau_{\rm reac}$ must be the largest time scale in order to observe a non-trivial $\Delta \sigma$. In the case of Fig. \ref{fig:Deltas}, the conductive timescale $\tau_{\rm cond}\simeq 3~$s, which corresponds nicely to the value of $\tau_{\rm reac}=(k^{\rm ads}\rho_{C,s})^{-1}$ that indicates the transition from a reaction-limited regime to another. Note that in the limit of $k^{\rm des}\rightarrow 0$ the surface flux $j_{\sigma}$ is limited by the slow exchange with the fluid, and its contribution to the net charge flux is negligible. This explains the small role played by the Duhkin number (Du) in the observed phenomena.

\end{document}